\documentclass[showpacs,twocolumn,amssymb,prl,aps]{revtex4}
\usepackage{graphicx,epsfig,verbatim,SIunits,color}

\newcommand{\etal}{\textit{et. al.}}

\begin{document}
\bibliographystyle{apsrev}
\title{Melting and freezing of argon in a granular packing of linear mesopore arrays}
\author{Christof~Schaefer, Tommy~Hofmann}
\author{Dirk~Wallacher}
\author{Patrick Huber}
\email[E-mail: ]{p.huber@physik.uni-saarland.de}
\author{Klaus Knorr}
\email[E-mail: ]{knorr@mx.uni-saarland.de}
\affiliation{Faculty of Physics and Mechatronics Engineering, Saarland University, D-66041 Saarbr\"ucken, Germany}

\date{\today}

\begin{abstract}
Freezing and melting of Ar condensed in a granular packing of template-grown arrays of linear mesopores (SBA-15, mean pore diameter $8~\nano\meter$) has been studied by specific heat measurements $C$ as a function of fractional filling of the pores. While interfacial melting leads to a single melting peak in $C$, homogeneous and heterogeneous freezing along with a delayering transition for partial fillings of the pores result in a complex freezing mechanism explainable only by a consideration of regular adsorption sites (in the cylindrical mesopores) and irregular adsorption sites (in niches of the rough external surfaces of the grains, and at points of mutual contact of the powder grains). The tensile pressure release upon reaching bulk liquid/vapor coexistence quantitatively accounts for an upward shift of the melting/freeezing temperature observed while overfilling the mesopores.
\end{abstract}


\maketitle
The arguably most conspicuous effect of pore confinement on molecular condensates is the shift of phase transitions with respect to the non-confined bulk state. Thus condensation (the vapor-liquid transition) does not occur at the saturated vapor pressure $p_0$ of the bulk liquid but in a wide range of reduced vapor pressures $P=p/p_0$, $P<1$, starting with the condensation of an adsorbed layer on the pore walls and culminating in pore filling via capillary condensation\cite{Cole1974}. In a similar way melting and freezing of the material in the pores occurs at temperatures $T$ well below the bulk triple point temperature $T_{\rm 3}$ \cite{Christenson2001}. These reductions are usually interpreted in terms of the Kelvin- and the Gibbs-Kelvin equation, resp. Here interfacial energies are introduced and because of their competition with volume free energies, the shift scales with the inverse of the characteristic linear dimension $L$ of the geometric configuration. In case of tubular pores this is the pore radius $R$.

Here we present an experimental calorimetric study on the melting and freezing of one of the most simple condensates imaginable, the rare gas Ar, in a template grown mesoporous silica, known as SBA-15 and considered one of the most ideal mesoporous substrates available \cite{Zhao1998}.
\begin{figure}[htbp]
\epsfig{file=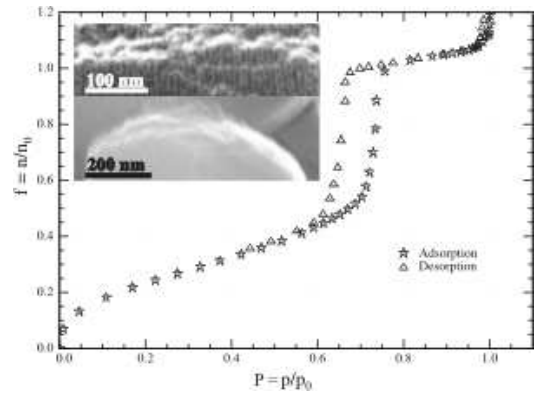, angle=0, width=0.8\columnwidth} \caption{\label{fig:ArIsotherme} Ar sorption isotherm in SBA-15, measured at $86~\kelvin$. Inset: Scanning electron micrographs of a SBA-15 grain taken with different spatial resolutions as indicated in the figure.}
\end{figure}

The preparation has been described elsewhere \cite{Hofmann2005}. The material obtained is a powder of grains, about $1$~micron in size, which have a relatively rough surface (Fig.~\ref{fig:ArIsotherme}). The pores within the grains are linear, non-ramified, parallel, have a relatively uniform cross section, and form a periodic 2D hexagonal array. Hence network effects and the blocking of the solidification front in bottle necks \cite{Khokhlov2007} are not expected to play a major role. The Bragg angles of the diffraction pattern of the empty pore lattice give a lattice parameter (= pore-pore-distance) of $10.7~\nano\meter$ \cite{Hofmann2005}. A fit of a radial electron density $\rho(r)$ profile to the Bragg intensities (with $\rho=0$ for $r<R_{0}-d$ , $\rho=\text{constant}$  for $r>R_{\rm 0}+d$, and with a linear function in the "corona" in between yields a pore radius $R_{\rm 0}$ of $3.8~\nano\meter$, corresponding to a porosity of $0.49$, and a corona thickness $2d$ of $2.2~\nano\meter$.
A volumetric Ar sorption isotherm (normalized uptake $f=n/n_0$ as function of the reduced vapour pressure $P$, where $n$ is the adsorbed amount of Ar and $n_0=30$~mmol is the amount of Ar for complete filling of the pores), recorded in the liquid regime of the Ar pore filling at $86~\kelvin$, is shown in Fig.~\ref{fig:ArIsotherme}. The initial reversible part is due to the condensation of Ar onto the pore walls, the hysteretic part to the filling of the pores by capillary condensation on $p$-increase (starting at about $f_{\rm A}=0.65$) and to the evaporation of the capillary condensate on $p$-decrease (terminating at about $f_{\rm D}=0.42$).  The branches of the hysteresis loop are relatively steep compared to what is observed for other mesoporous substrates, suggesting a narrow distribution of the pore radius and the absence of pore network effects such as pore blocking \cite{Yortsos1999}. The final slope in the post-filling regime $f>1$ (absent e.g. for monolithic Vycor substrates) demonstrates that there are further sites available for condensation, with characteristic linear dimensions larger than $R_{\rm 0}$, but still finite.  We think of tapered pore mouths, niches of the rough external surface of the grains, and the points at which the powder grains are in mutual contact.

\begin{figure}[htbp]
\epsfig{file=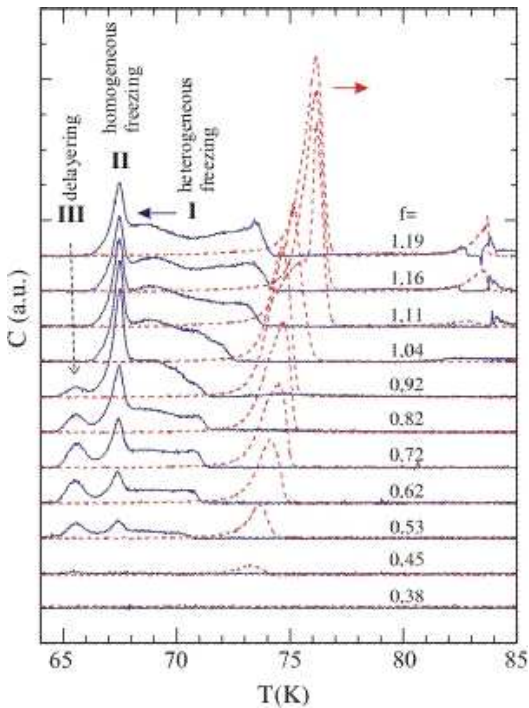, angle=0, width=0.8\columnwidth} \caption{\label{fig:c_heatcool}(color online). Specific heat curves of cooling (solid line) and heating (dashed line) cycles for selected filling fractions $f$ indicated in the figure.}
\end{figure}

The calorimeter consists of the sample cell, heat shield, and the outer jacket, with two separate vacuum spaces in between. The sample cell and the heat shield are equipped with thermometers ($1000~\Omega$ Pt-resistors) and resistance heaters. The heat shield is in thermal contact with the cold plate of a closed cycle He refrigerator.

The cell contains $1.3~\gram$ SBA-15 and a coil of Ag wire (that is meant to reduce the thermal time constant within the powder). $12$ fractional fillings have been investigated that have been prepared by introducing suitable amounts of Ar into the cell via a fill line at $86~\kelvin$.

We employ a scanning experiment the calorimetric signal is proportional to $dQ/dT$,the amount of heat per unit $T$-interval that goes into or comes out of the sample. $dQ/dT = K \Delta T/r$, $\Delta T$ and $K$ are the temperature difference and the heat leak between cell and shield, $r$ is the cooling/heating rate of the shield. The heat leak has been realized by a He gas pressure of a few mbar in the vacuum space between cell and shield, $\Delta T$ was typically $1~\kelvin$, the heating/cooling rate was held constant at a value of $0.05~\kelvin\per\minute$. Quite elaborate thermometry allowed us to work with such low rates which are more than one order of magnitude smaller than what is standard with commercial equipment.

The samples have been subject to a complete freezing/heating cycle starting at $86~\kelvin$ in the liquid regime, taking the sample to a minimum temperature of $50~\kelvin$ deep in the solid regime, and bringing it back to $86~\kelvin$. The results for a selected set of fillings are shown separately for cooling and heating in Fig.~\ref{fig:c_heatcool}.

Samples with $f\geq 0.45$ do show heating and freezing anomalies, samples with $f\leq 0.38$ do not. Analogous observations have been made for Ar in Vycor \cite{Wallacher2001}. Thus a phase-transition like change from the liquid to the crystalline state is basically reserved to the capillary condensed part of the pore filling. The initial "dead" fraction of the condensate up to about $f=0.4$, adsorbed on the pore walls and filling the niches of the corona, does not participate in a collective freezing melting process.

\begin{figure}[htbp]
\epsfig{file=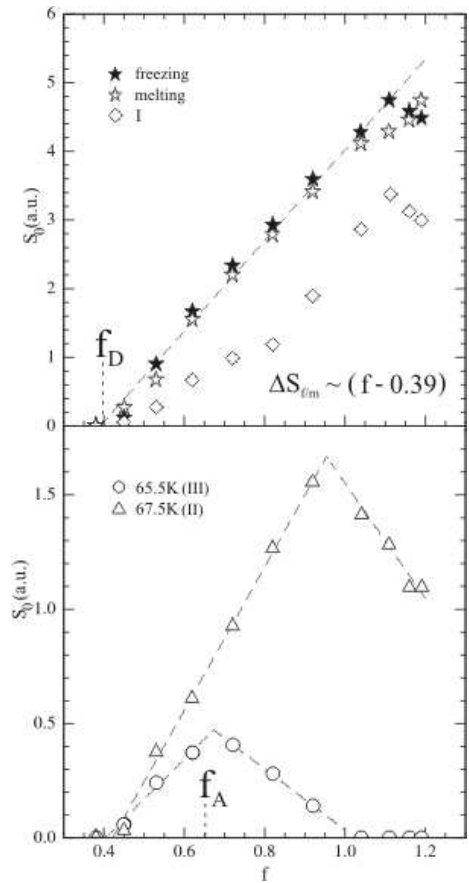, angle=0, width=0.7\columnwidth} \caption{\label{fig:S}Entropy plots of the different anomaly parts as a function of filling fraction $f$. The dashed lines are guides for the eye.}
\end{figure}

Melting of the "mobile" part occurs in a relatively narrow $T$-interval of $2~\kelvin$, which for $f<1$ is centred at $74.5~\kelvin$, whereas freezing - say for $f=0.92$ - extends over $7~\kelvin$ from $71.5~\kelvin$ down to $64.5~\kelvin$ with the freezing anomaly having a rather complex shape (Fig. \ref{fig:c_heatcool}). The  entropy of fusion, $S_{\rm 0}$, is obtained by integrating the calorimetric signal above the smooth background, $S_{\rm 0} = c \smallint(dQ'/dT)/TdT$. The coefficient $c$ has not been determined, but is the same for all heating and cooling runs. A plot of $S_{\rm 0}$ as function of $f$ is shown in Fig.~\ref{fig:S}. The values of $S_{\rm 0}$ obtained from cooling and heating runs are identical within experimental error. This is not only a consistency check but also means that all parts of the freezing anomalies are indeed due to no other phase transition but solidification. For $f<1$, $S_{\rm 0}$ is a linear function of $f$ that extrapolates to zero at about $f=0.39$, close to $f_{\rm D}$ of the $86~\kelvin$-sorption isotherm.  Thus the part of the condensate that supports the collective freezing/melting process comprises not only what is formed by capillary condensation but also the fraction which grows initially in form of a metastabel film at the pore walls ($f_{\rm A}-f_{\rm D}$).

For ease of discussion we divide the freezing anomaly into the parts I,II,III which are meant to pertain to certain fractions of the pore filling. Part~II and III are the peaks at $67.5~\kelvin$ and $65.5~\kelvin$, respectively, part~I the remainder at higher $T$. The $f$-dependence of the three parts is also shown in Fig.~\ref{fig:S}.

The fact that the pore filling solidifies sequentially, but melts practically in a single step strongly suggests that solidification leads to a reorganization of the capillary condensate upon which the different fractions I, II, III of pore material loose their identity.

The melting anomalies have an asymmetric shape. For $f<1$, the low-$T$ wings of the anomalies for different filling fractions superimpose. We have interpreted the analogous observations on Ar in Vycor in terms of interfacial melting in a cylindrical pore starting at the boundary between the dead and the mobile part of the pore filling rather than by referring to a pore size distribution which had to be rather special in order to reproduce the peculiar shape of the anomalies \cite{Wallacher2001}. This interfacial melting scenario is similar to surface melting at semi-confined, planar surfaces \cite{Dash1986}. Calorimetry can of course not prove whether the melting front propagates along the pores or radially inward, but the mere fact that the asymmetric shape of the anomaly is recovered in the highly homogeneous pores of the present substrate is a strong argument in favor of interfacial i.e. radial melting.

As for freezing, part~III of the freezing anomaly represents the component of the filling that is the last to solidify on cooling. $S_{\rm 0}$(III) shows the peculiar type of $f$-dependence , $S_{\rm 0}$(III) scales with the vapor filled fraction of pore space. $S_{\rm 0}$(III) is due to the solidification of the mobile part of the liquid film on the pore walls in otherwise empty pore segments. Upon solidification the film delayers and the material involved joins the already solidified capillary condensate (and melts as such on subsequent heating). The same observations have been made for Ar in Vycor \cite{Wallacher2001}.

Still referring to the underfilled situation, $f<1$, this reasoning leaves parts I and II for the freezing of the capillary condensate in the regular pores. We recall that the interfacial melting model not only explains the lowering of the equilibrium freezing/melting temperature $T_{\rm 0}$ in pore confinement but also predicts metastable states, liquid ones down to the lower spinodal temperature $T^{\rm -}$ and solid ones up to the upper spinodal temperature $T^{\rm +}$. We propose to identify peak~II with the freezing of the capillary condensate via homogeneous nucleation of the solid phase at $T^{\rm -}$. Peak~II is sharp because of the narrow pore size distribution of SBA-15. Part~I of the freezing anomaly on the other hand stems from the freezing via heterogeneous nucleation within the rather broad ($T_{\rm 0}$, $T^{\rm -}$)-interval, triggered by the presence of various types of small nuclei residing at irregular sites. The chance that a parcel of liquid is in contact with such nuclei increases with $f$. That is why part~II dominates (in coexistence with part~III) at low $f$ and goes through a maximum slightly below $f=1$ whereas part~I gradually develops with increasing $f$ in a somewhat delayed fashion (Figs.~\ref{fig:c_heatcool} and \ref{fig:S}).

The data on the overfilled sample, $1.04<f<1.19$, shows that the freezing/melting of different parts of the Ar condensate are not independent from each other, an observation that supports the idea of heterogeneous freezing. Quite in contrast to the results on a monolithic Vycor sample, the melting and freezing of the excess material, $f-1$, does not lead to a $\delta$-peak at the triple point of the bulk system (at $83.8~\kelvin$), but in first respect to some extra calorimetric signal at the high-$T$ end of the anomalies, both for freezing and melting. Obviously most of the excess material resides in sites, say in niches of the external surface of the powder grains, with dimensions that are still in the nm-range just slightly larger than the diameter of the regular pores. The presence of this material has an effect on the freezing and melting of the material in the pores. The melting peak for $f=0.92$ for example is centered at $T_{\rm m}=74.8~\kelvin$, but adding excess material the melting peak eventually shifts to $76.1~\kelvin$ and even reduces the calorimetric signal at the former $T_{\rm m}$. Thus the presence of material outside stabilizes the solid state in the pores.

The pore filling experiences a negative hydrostatic pressure $p_{\rm h}$ of $70\pm10$~bar for $f<1$ \cite{Kanda2004}. This value is dictated by the Laplace equation while assuming semi-spherical, concave menisci with negative curvature radii of the condensate-vapor interface on the order of the pore radius \cite{Huber1999}. Upon post-filling of the pores the negative curvature of these interfaces gradually vanishes; hence the tensile pressure in the pore condensate disappears. Based on the $p_{\rm h}$ dependency of the melting line of Ar \cite{Michels1962}, an increase of the melting temperature of $1.7\pm0.3$~K is expected for a $70\pm10$ pressure release, in good agreement with the $\sim$ 1.5~K $T$-shift of the melting peak documented in Fig.~2. The upward shift of the onset of the freezing anomaly for $f>1$ (part I) can be explained in an analogous way. An opposite, downward shift of $T_{\rm m}$ upon approaching $f=1$, observed for water in mesopores \cite{Findenegg2001}, is consistent with our tensile pressure hypothesis due to water's anomalous $T_{\rm m}(p_{\rm h})$-curve.
\begin{figure}[htbp]
\epsfig{file=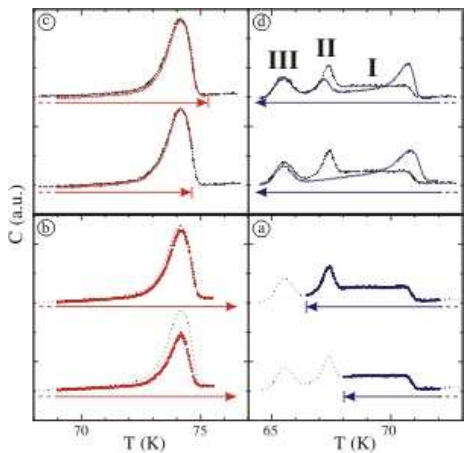, angle=0, width=0.7\columnwidth} \caption{\label{fig:partial}(color online). Incomplete (partial) cooling-heating cycles at $f=0.66$ (solid lines) in comparison to heating, cooling curves after complete solidification, melting, resp. at $f=0.62$ (dotted lines).}
\end{figure}

For $f=0.66$ partial cooling/heating cycles have been investigated (Fig.~\ref{fig:partial}). In the first type of such cycles a part of the condensate is solidified by cooling to some temperature within the $T$-range of the freezing (Fig.~\ref{fig:partial}a) and the melting of this part is then studied on subsequent heating (Fig.~\ref{fig:partial}b). In the first run the freezing of part I is initiated by cooling down to $68~\kelvin$, in the second one parts~I and II are solidified by cooling down to $66.5~\kelvin$. As far as position and shape are concerned, the resulting melting anomalies are identical to what is obtained after complete solidification (see Fig.~\ref{fig:c_heatcool} for data on $f=0.62$ and $0.72$), it is just that $S_{\rm 0}$ is scaled down in proportion to the amount of material solidified on cooling. Parts of the condensate with largely different freezing temperatures melt in the same narrow $T$-interval.

The second type of cycles starts at $60~\kelvin$ deep in the solid regime and a part of the solid is melted by heating up to temperatures within the $T$-range of melting (Fig.~\ref{fig:partial}c). The re-solidification is then studied by subsequent cooling (Fig.~\ref{fig:partial}d). In such partial cycles the calorimetric signal right at the onset of freezing is enhanced. Obviously the fraction of material that has been melted in the heating cycle is still in contact with the solid remaining and can therefore resolidify directly without having to overcome a nucleation barrier. Peak~III is recovered in the partial cycles. Whenever some part of the pore filling is liquefied, the mobile part of film coating on the pore walls is re-established.  As for peak~II, the completed freezing process of the first cooling run down to 60~K leads to a re-arrangement of the pore filling in pore space by which the isolated parcels are eliminated, the freezing of which (via homogeneous nucleation) gives rise to peak~II. These parcels are not re-established in case melting along the heating run is incomplete. This is why peak~II is absent in the lower trace of Fig.~4d. Even in case of complete melting by heating up to a temperature of 75.5~K slightly above the high-$T$ cutoff of the melting anomaly, the isolated parcels are not fully recovered compared to the situation of the virgin sample right after preparation by vapor condensation at 86~K, see the dotted line in Fig.~4. This explains the reduced size of peak~II in the upper trace of Fig.~4d. After a freezing-melting cycle the pore liquid keeps some memory of the spatial distribution of the pore solid \cite{Soprunyuk2003}.

The freezing/melting phenomenology encountered in SBA-15 challenges the common notion which underlies pore size spectroscopy techniques (''thermoporometry'') \cite{Brun1977} relying on simple geometric relations between pore diameter and specific heat anomalies. As demonstrated here for a prototype mesoporous system, neither a resorting to the specific heat anomaly in freezing nor in melting will help in a granular packing of mesoporous grains. Freezing is demonstrated to be significantly affected by heterogeneous nucleation, not to mention the delayering transition for partial fillings, whereas melting is governed by interfacial melting processes.

Freezing and melting in such a granular packing of comparably simple linear mesopores is clearly a complicated process. Metastable states, an interplay of homogeneous and heterogeneous nucleation processes and the coupling between different regions of pore space lead to a remarkable complex phenomenology.







\begin{acknowledgments}
This work has been supported by the SFB 277 of the Deutsche Forschungsgemeinschaft.
\end{acknowledgments}

\end{document}